\documentstyle[prl,twocolumn,aps]{revtex}

\begin{document}
\bibliographystyle{unsrt}
\input{psfig}

\twocolumn[{
\title{
Do Topological Charge Solitons Participate
in DNA Activity?}
\author {Z. Hermon$^1$, S. Caspi$^2$ and E. Ben-Jacob$^2$}
\address{
$^1$Institut f\"ur Theoretische Festk\"orperphysik,
Universit\"at Karlsruhe, D-76128 Karlsruhe, Germany. \\
$^2$ School of Physics and Astronomy, Tel Aviv University,
69978 Tel Aviv, Israel.}
\maketitle
\date{\today}
\vspace*{-.4in}

\begin{abstract}
\widetext
\leftskip 54.8 pt
\rightskip 54.8 pt

We present a novel electromagnetic model of DNA molecules, in which
the P-bonds act as tunnel junctions and the H-bonds as capacitors.
Excess charge in the model gives rise to two coupled modified 
sine-Gordon equations, which admit topological solitonic excitations.
We study the dynamics of the solitons, their effect on the DNA
transport properties, and comment about their role in the DNA functioning. 
We propose specific experiments in order to test our predictions.
\end{abstract}
}]
\narrowtext

The DNA molecule is central to every living organism. It carries
the genetic code of the organism, and continuously controls the 
synthesis of proteins, which are vital to its functioning and existence. 
Since the pioneering discoveries of Watson and Crick, a considerable
effort was devoted to the study of the DNA, and much has been understood
about its structure and activity \cite{Biochemistry}. Yet there are some key 
features which pose open questions. One such feature is the long range 
correlation and control between segments of the DNA. Synthesis 
of proteins is done via local ''reading'' of specific sectors in the DNA. The 
initiation of the reading is done by another segment, which can be located
far away along the DNA sequence (thousands of bases away). Consider this 
and other examples \cite{Genome_Wisdom}, it seems as if the DNA molecule 
has the capability of transmitting information over long distances and 
in a specific manner (the information is transmitted to/from specific 
targets). Possible efficient candidates for such a transmission are solitons
and solitary waves (rather than point particles/wave packets and linear 
waves). The existence and propagation of conformational solitons related to
the DNA replication process have been studied in the past \cite{Replic_sol}. 
Non-topological charge solitons in proteins have also been studied 
\cite{Davidov}. Motivated by our studies of topological charge solitons in 
one dimensional (1D) arrays of mesoscopic tunnel junctions 
\cite{Normal_Solitons}, \cite{Charge_Soliton}, we have investigated the 
possibility of the existence of topological charge solitons in DNA. 

\begin{figure}
\centerline{
   \hbox{
      \psfig{figure=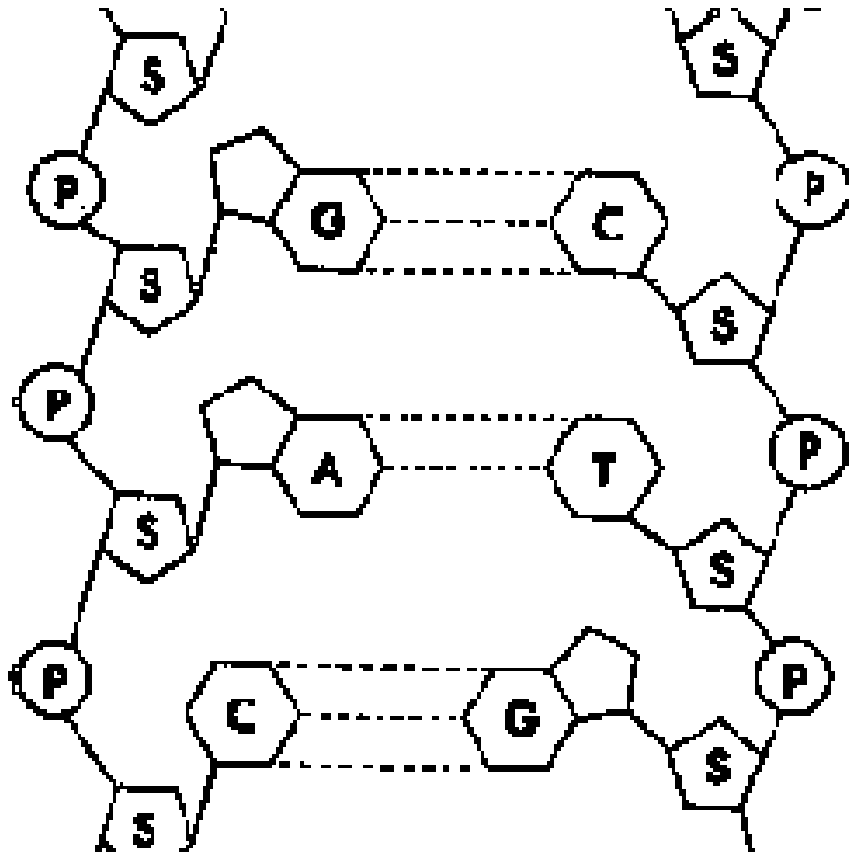,height=1.4in}
      \hspace{0.2in}
      \psfig{figure=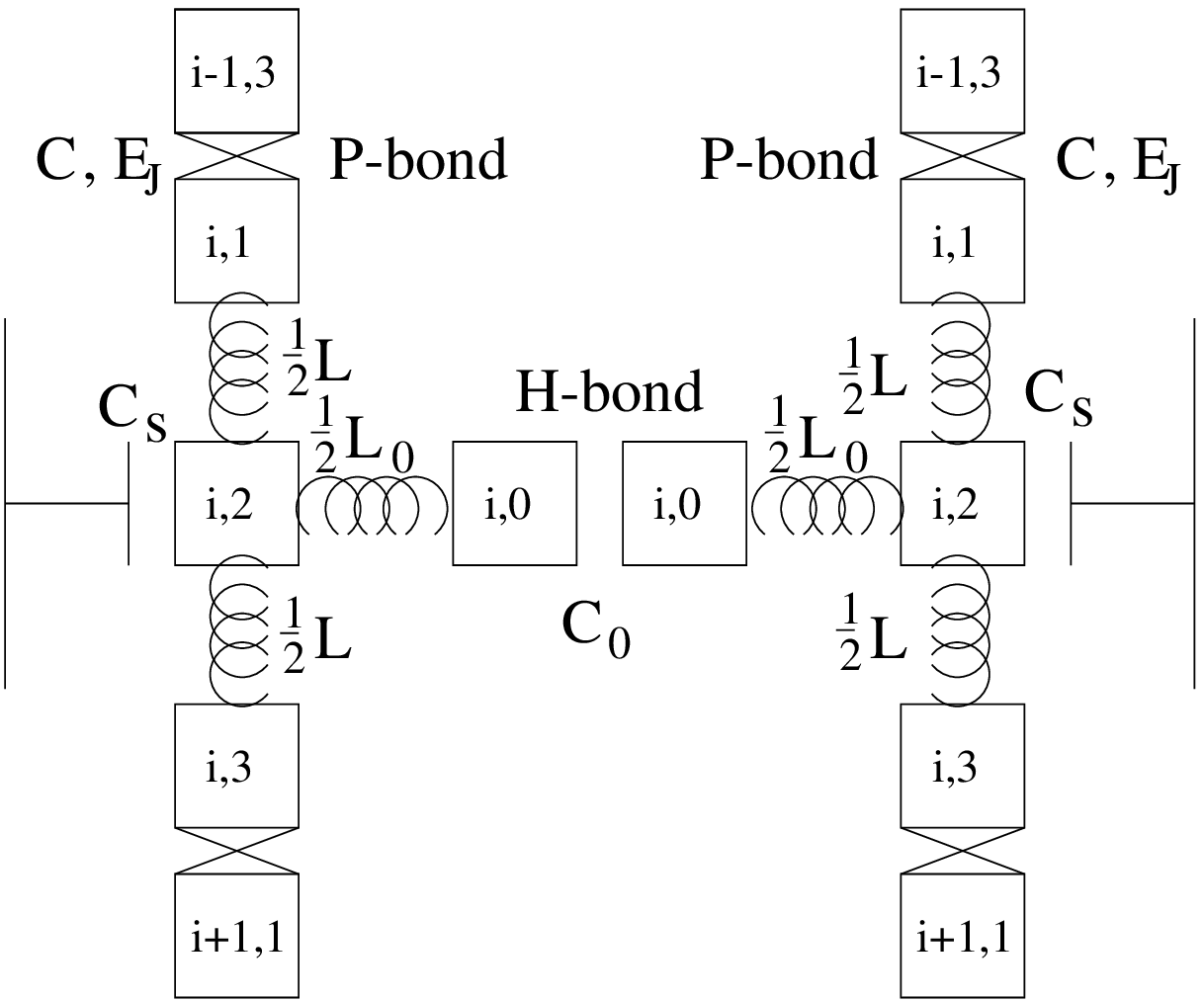,height=1.4in}
   }
}
\caption{A schematic image of a DNA molecule (left), and the double-strand
model (right). On the left, P denote the P-bonds between the sugars S and the 
dashed lines are the H-bonds between the bases A,G,T and C. The parameters 
of the double-strand model are described in the text. In the single-strand 
case only one half of the model is used, and $C_0$ denotes the capacitance 
to the outside world.}
\label{DNA_fig}
\end{figure}

We have developed a novel electromagnetical model of the DNA molecule,
which is based on the properties of charge dynamics in the DNA (see 
Fig.~\ref{DNA_fig}). We view each 
of the DNA strands as a 1D array. A unit (or 'grain') of the array is a sugar 
and a base attached to it. The grains are connected longitudinally by 
Phosphate groups (P-bonds), and parallely, i.e., to the other strand, by the 
Hydrogen bonds (H-bonds) between bases. An additional electron residing on an 
atom belonging to the sugar-base grain can hop from atom to atom, thus 
obtaining a kinetic energy. The H- and P-bonds 
form barriers to the charge propagation. The proton in the H-bond can 
effectively screen a net charge density on either side of the bond by shifting 
its position towards this side. (By 'net' charge we mean the change from the 
charge distribution of the unperturbed DNA.) As a result, electrons do not 
cross the H-bond. Hence the bond can be viewed as a capacitor. The P-bond 
barrier stems from the two oxygens which are transversely connected to the 
phosphorus. These oxygens share three electrons with the phosphorus, giving 
rise to two $\sigma$ bonds and one $\pi$ bond. As the $\pi$ electron can be 
shared with both oxygens, it behaves as an electron in a double well, and 
occupies the lowest level. When another electron approaches the well it 
encounters a barrier due to the energy difference to the next level of the 
well. However, since this barrier is narrow, the approaching electron can 
tunnel through the well. Thus from the charge dynamics point of view, the 
P-bond behaves as a tunnel junction.

We start with the model for a single strand (see Fig.~\ref{DNA_fig}). 
For simplicity we assume that all the bases are of the same type. Each grain 
$i$ is composed of four sub-grains ($j=0,1,2,3$), assigned with phase 
variables, $\phi_{i,j}$. These phases are related to the electric potential of 
the sub-grains through 
$\phi_{i,j}(t)\equiv{1\over\hbar}\int_{-\infty}^t\,V_{i,j}(t')dt'$. The 
conjugate variable to each $\phi_{i,j}$ is the charge on the sub-grain, 
$Q_{i,j}$. The Lagrangian of this model is
\begin{eqnarray}
\label{Full_Lagrangian_Single}
L&=&\sum_i \left[{C\over 2}(\dot\phi_{i+1,1}-\dot\phi_{i,3})^2+
{C_S+C_0\over 2}\dot\phi_{i,2}^2+
\right.
\nonumber \\
& & {1\over 2L/2}(\phi_{i,1}-\phi_{i,2})^2-
{1\over 2L/2}(\phi_{i,2}-\phi_{i,3})^2+
\nonumber \\
& & \left. E_J\cos(\phi_{i+1,1}-\phi_{i,3})-
{1\over L_0/2}\left(\phi_{i,0}-\phi_{i,2}\right)^2 \right]\ .
\end{eqnarray}
It includes three types of energies: inductive energies which represent the
hopping of electrons, capacitive energies which represent the capacitive 
properties of the H- and P-bonds, and a tunneling energy, which represents the
tunneling process in the P-bond. This latter energy is proportional to the
cosine of the phase difference across the bond, according to the tight binding
picture \cite{Tight_Binding}. The values of the parameters can be obtained, 
in principle, from experiments, but for now we employ only a qualitative view. 
The $L_0/2$ inductance denotes the hopping from the open side of a base to
its sugar, while the $L/2$ inductance denotes the hopping between P-bonds.
Since the former involves more hopping around carbon rings, we assume that 
$L_0>L$. Both $C_S$ and $C_0$ denote in this single-strand model
capacitances to the outside world. $C_S$ is the capacitance seen from the
sugar, while $C_0$ is the capacitance seen from the open H-bond. We assume that
both are of the same order of magnitude and much smaller than $C$, the
capacitance of the P-bond tunnel junction. $E_J$ is the tunneling strength of
the P-bond. The length scale in our model, $a$, is the 
distance between grains, which is $3.4\AA$. (In the model we use units in 
which $a=1$.) 

{}From the kinetic part of (\ref{Full_Lagrangian_Single}) we see that out 
of the four variables describing the $i$'th grain only three are independent. 
According to our experience in the study of 1D arrays of tunnel junctions
\cite{Charge_Soliton}, we introduce non-local (or integral) charge variables.
These variables are very useful in trying to probe the collective, non-local
dynamics of the DNA. We envision a propagation of charge from a certain
point in the far left of the chain ('$-\infty$') to a certain point in the far
right ('$\infty$'), passing on its way the $i$'th grain. Referring
to the $i$'th junction as the junction between the $i$'th and the $(i+1)$'th
grains, we define $R_i$ as the charge that has {\it passed} through this
junction:
\begin{eqnarray}
\label{R_Variable_Single}
R_i & \equiv & \sum_{i+1,1}^{\infty}Q_{i',j}  =  \sum_{i+1}
(Q_{i',1}+Q_{i',2}+Q_{i',0}+Q_{i'})-Q_{\infty}
\nonumber \\
& & =  Q_{i,1}+\sum_{i+1}(Q_{i',2}+Q_{i',0}) \ .
\end{eqnarray}
$\sum_{i+1,1}^{\infty}Q_{i',j}$ means summation over the sub-grain charges,
starting from $(i'=i+1,j=1)$ and going right, with the internal order $(i+1,1)
\rightarrow(i+1,2)\rightarrow(i+1,0)\rightarrow(i+1,3)\rightarrow(i+2,1)$,
and so on. Next we define $q_i$ as the charge that has {\it reached}
junction $i$:
\begin{equation}
\label{q_Variable_Single}
q_i \equiv \sum_{i,3}^{\infty}Q_{i',j} =  Q_{i,3}+R_i=
\sum_{i+1}(Q_{i',2}+Q_{i',0}) \ .
\end{equation}
After the canonical transformation to the variables: 
$\pi_i\equiv\phi_{i+1,2}-\phi_{i,2}-\theta_i$, $\theta_i\equiv 
\phi_{i+1,1}-\phi_{i,3}$
and $\tilde\phi_{i,0}\equiv\phi_{i,0}-\phi_{i,2}$,
we obtain the following Euclidean Lagrangian (in the continuum limit 
which we justify later):
\begin{eqnarray}
\label{Euclidean_Lagrangian_Single}
L_E&=&\int\,dx\left\{{1\over 2}L\dot q^2(x)-E_J\cos\theta(x)+
{1\over 2}{L_0\over 2}\dot q_0^2(x)+
\right. \nonumber \\ 
& & {1\over 2C}\left[q(x)-R(x)\right]^2+
{1\over 2C_S}q_x^2(x)+
\nonumber \\
& & \left. {1\over C_S}q_0(x)q_x(x)+{1\over 2C_{S0}}q_0^2(x)+
i\dot\theta(x)R(x) \right\} \ ,
\end{eqnarray}
where $C_{S0}\equiv C_SC_0/(C_0+C_S)$ is the effective capacitance across
the chain (parallel coupling of $C_S$ and $C_0$). As $C_S$ is much
smaller than $C_0$, $C_{S0}\approx C_S$. We define a tunneling inductance
$L_J\equiv \Phi_0^2/(2\pi E_J)^2$ ($\Phi_0\equiv h/e$ is the flux quantum),
which we take to be much smaller than $L$. {}From 
(\ref{Euclidean_Lagrangian_Single})
one can identify three characteristic frequencies in the system:
$\Omega^2\equiv 1/(LC)$ of the $q$ mode,  
$\omega_J^2\equiv 1/(L_JC)$ of the $R$ mode, and 
$\omega^2\equiv 2/(L_0C_{S0})$ of the $q_0$ mode. 
In the limits we are working, we have $\Omega^2\ll\omega_J^2, \omega^2$.
Thus we can integrate out the $R$ and $q_0$ modes, and obtain an effective
theory for the $q$ mode. The effective real time Lagrangian is
\begin{eqnarray}
\label{Full_eff_Lagrangian_Single}
L^{eff}&=&\int\,dx\,\left\{{1\over 2}L\dot q^2-{2\over(2\pi)^2}E_C
\left[1-\cos\left({2\pi\over e}q\right)\right]-
\right.  \nonumber \\
& & \left. {1\over 2(C_0+C_S)}q_x^2+
{1\over 2}{C_0^2L_0\over 2(C_0+C_S)^2}\dot q_x^2\right\}\ ,
\end{eqnarray}
where $E_C\equiv e^2/(2C)$. The equation of motion of 
(\ref{Full_eff_Lagrangian_Single}) is the following modified sine-Gordon 
equation:
\begin{eqnarray}
\label{Single_Eq}
L\ddot q+V_D\sin\left({2\pi\over e}q\right)-{1\over C_0+C_S}q_{xx}-
{C_0^2L_0\over 2(C_0+C_S)^2}\ddot q_{xx}
\nonumber \\ =0 \ ,
\end{eqnarray}
where $V_D\equiv e/(2\pi C)$. This is a Kirchoff law for the equivalent
electrical circuit. Since the characteristic length scale in a sine-Gordon
model is $C/(C_0+C_S)$, the continuum limit is justified when $C>C_0+C_S$,
which indeed corresponds to our assumption.
Transforming into dimensionless space, time and charge variables:
$x\rightarrow x'\equiv\sqrt{C/C_0}x, \ \ t\rightarrow t'\equiv\sqrt{LC}t, \ \ 
q\rightarrow q'\equiv {e\over 2\pi}q$, we obtain:
\begin{equation}
\label{Scaled_Single_Eq}
\ddot q+\sin q-{1\over 1+\mu_S}q_{xx}-{\zeta\over 2(1+\mu_S)^2}
\ddot q_{xx}=0\ ,
\end{equation}
where $\zeta\equiv{L_0C_0\over LC}$, and $\mu_S\equiv C_S/C_0$.

Next we introduce the double-strand model (see again Fig.~\ref{DNA_fig}). 
We distinguish between the two strands by the superscripts $\alpha, \beta$. 
The capacitance $C_0$ describes now the capacitive coupling between the two 
strands, i.e., over the H-bonds. We assume that it is much larger than the 
capacitance to the outside world, $C_S$, but still smaller than $C$. We take 
the parameters of the two strands to be equal. Following the same steps of 
derivation as in the single-strand model (details will be given elsewhere 
\cite{Next_paper}), we obtain two (dimensionless) coupled modified sine-Gordon 
equations of motion:
\begin{equation}
\label{AVE2_Eq}
\ddot{\bar q}+\sin\bar q\cos q-{1\over 2\mu}\bar q_{xx}=0 \ ,
\end{equation}
\begin{eqnarray}
\label{REL2_Eq}
\ddot q+\sin q\cos\bar q-{1\over 2(1+\mu)}q_{xx}-
{\zeta\over 2(1+\mu)^2}\ddot q_{xx}=0 \ ,
\end{eqnarray}
where $\mu \equiv C_S/2C_0$, and we have used the average and the relative 
charge variables:
\begin{equation}
\label{New_Charges}
\bar q\equiv{1\over 2}\left(q^\alpha+q^\beta\right) \hskip 2cm 
q\equiv{1\over 2}\left(q^\alpha-q^\beta\right) \ ,
\end{equation}
Similar equations, without the $\zeta$ term, where studied
in the past in connection with the stacked Josephson junctions model 
\cite{Coupled_Josephson1}.

We turn now to study the modified sine-Gordon equations we have derived,
starting from the single-strand model. The pure Sine-Gordon equation has 
exact topological solitons solutions. We have checked numerically and found 
that the extra term in equation (\ref{Scaled_Single_Eq})
does not effect the stability of the topological soliton solution, though
it does induce interaction with the plasmons. This interaction causes the 
soliton to slowly radiate away its kinetic energy. Using the collective 
coordinate:
\begin{equation}
\label{Collective_coord_eq}
  X = -{1 \over 2\pi}\int q_x x\,dx,
\end{equation}
we find that the soliton velocity actually oscillates with relatively high 
frequency, but its average is almost constant, decreasing only slightly over 
large period of time. Therefore, though the soliton's motion is not persistent 
in the exact sense, charge can propagate relatively large distances along the 
strand without the need to apply a driving force in the form of a potential 
difference.

The extra term in equation (\ref{Scaled_Single_Eq}) destroys the Lorenz 
invariance of this equation. The dispersion relation for small amplitude 
linear waves is
\begin{equation}
  \omega^2 = {k^2/(1+\mu_S) + 1\over {\zeta\over 2}k^2/(1+\mu_S) + 1}\ ,
\end{equation}
and the group velocity tend to zero for both small and large k values.
When the soliton velocity exceed the maximal group velocity, it leaves in its 
wake all the small amplitude waves.

In the double-strand model, the dimensionless parameter $\mu$ serves as a 
measure of the strength of the interaction between the strands (it is small 
for strong interaction). Since $C_S \ll C_0$, $\mu$ is very small. In the 
extreme $\mu = 0$ limit, $\bar q_x$ should be zero as well, in order that 
the Lagrangian for this model would be finite. As both $q^\alpha$ and $q^\beta$ 
have integer values at the two edges of the chain, $\bar q\approx 2\pi n$ 
everywhere. We are thus left with a single equation for q:
\begin{equation}
\label{Double_q_Eq}
\ddot q+\sin q-{1\over 2}q_{xx}-{\zeta\over 2}\ddot q_{xx}=0 \ .
\end{equation}
This equation is equivalent to the one of the single-strand model, but now $q$
represents a soliton anti-soliton pair (with zero total charge) rather than
a single charged soliton.

Equations (\ref{AVE2_Eq},\ref{REL2_Eq}) have two simple topological solutions. 
One is the "symmetrical" solution, namely, $q=0$ and $\bar q$ is a usual 
Sine-Gordon kink. This solution is unstable ,at least at low energy, since it 
corresponds to two charges with equal sign which tend to separate. It is, 
however, possible that it becomes stable at high energy, as was demonstrated 
for the stacked Josephson junctions model \cite{Coupled_Josephson2}. The other 
"anti-symmetrical" solution has $\bar q=0$ and q in the form of a kink. It is a 
stable solution which correspond to an electron-hole pair (each on a different 
strand). It is plausible that this excitation is used in the DNA to transmit 
information over long distances. It can be created at a specific segment, which 
is responsible for the initiation of a certain protein synthesis, by tunneling 
of an electron through the H-bond (as a result of external agent). The pair 
can, then, propagate almost freely along the DNA, and be annihilated at the
protein synthesis segment by a reverse tunneling, thus transmitting the order
for this synthesis. The targeting of a specific annihilation segment is
probably done through the specific base sequence of both segments, which was
neglected in our model.

\begin{figure}
\centerline{\psfig{figure=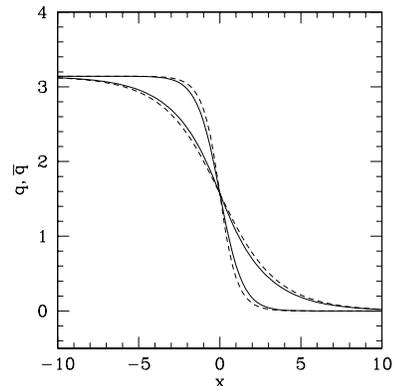,height=2.1in}}
\caption{Topological solution representing stationary charge in the $\alpha$
strand. The values $\mu = 0.1$ and $\zeta = 1$ where used. The steepest curve 
is $q(x)$, and the other is ${\bar q}(x)$. Numerical results are indicated by 
continues lines and the theoretical curves of equation
({\protect \ref{aprox_sol_eq}}), by dashed lines.}
\label{soliton_fig}
\end{figure}

There is also a way to approximate a solution which corresponds to an electron 
(hole) moving with velocity $v$ in the, say, $\alpha$ strand:
\begin{eqnarray}
\label{aprox_sol_eq}
  q & = & 2 \arctan\,\exp
                     \left[ -\gamma(x - vt)/c \right] \nonumber \\
  \bar q & = & 2 \arctan\,\exp
                     \left[ -\bar\gamma (x - vt)/\bar c \right],
\end{eqnarray}
where $\gamma$ is the relativistic factor, and $c=1/\sqrt{2(\mu + 1)}$,
$\bar c=1/\sqrt{2\mu}$, are the linear wave velocities in equations 
(\ref{AVE2_Eq},\ref{REL2_Eq}) (when the last term in equation (\ref{REL2_Eq}) 
is neglected). They correspond to the different linear waves velocity in the 
two equations. The exact solution can be obtained numerically and is shown in 
Fig.~\ref{soliton_fig}.

\begin{figure}
\centerline{\psfig{figure=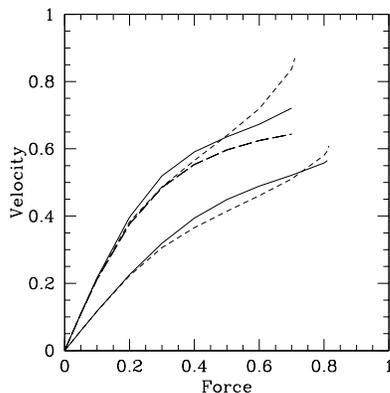,height=2.1in}}
\caption{The numerical current (velocity) - voltage (force) curves for a charge 
soliton propagating in one strand. In the lower curves $\alpha = 0.5$ and 
$\mu = 0.5$, while in the upper ones $\alpha = 0.5$ and $\mu = 0.05$. In the
continuous lines $\zeta = 0.1$, and in the short dash lines $\zeta = 1$. The 
theoretical curve of equation ({\protect \ref{Theoretic_IV_eq}}) is indicated 
by a long dash line.} 
\label{IV_fig}
\end{figure}

An important way to test our model is via measurements of the current-voltage 
characteristics. An electron can be injected to one of the strands to form a 
topological charge soliton, and may be subjected to an applied external
voltage $V$ and an Ohmic dissipation $R$ (both per unit length). We implement 
this in our model by adding the terms $F + \alpha \dot{\bar q}$ to equation 
(\ref{AVE2_Eq}) and $\alpha \dot{\bar q}$ to equation (\ref{REL2_Eq}), where
$F \equiv {V/ V_D}$ and $\alpha \equiv {RC/\sqrt{LC}}$, are the external force 
and the dissipation rate respectively. For given values of force and 
dissipation rate a soliton reaches a limiting velocity which corresponds to the 
measured current. The I-V curves which we obtain for different choices of
the $\mu$ and $\zeta$ parameters are shown in Fig.~\ref{IV_fig}. It should be 
noted that above some threshold voltage close to the maximum voltage indicated 
in each curve of Fig.~\ref{IV_fig}, the system becomes unstable, as more and 
more soliton anti-soliton pairs are created.

We proceed to estimate the limiting velocity. To do so, we equate the energy 
gain by the external force and the dissipative energy \cite{Power_balance}. We 
find that $\dot{X} (\gamma/c + \bar\gamma/\bar c) = {\pi F/2\alpha}$.
This reduces in the small $\mu$ limit to
\begin{equation}
\label{Theoretic_IV_eq}
  \dot{X}={1\over\sqrt{2\left[\left({2\alpha\over\pi F}\right)^2 + 1\right]}}.
\end{equation}
This theoretical curve is also shown in Fig.~\ref{IV_fig}. The ascending of 
the numerical curve in the ``relativistic'' regime above the theoretical curve 
is obviously due to the $\zeta$ term.\

We turn now to discuss some possible experimental tests of our predictions. 
The most straightforward test is to measure the I-V characteristics of a DNA 
that was mentioned earlier. Such experiments are now being developed 
\cite{Experiment}. We propose to use first a single strand vs a double strand. 
It is easier to inject solitons into the former. In the case of a double 
strand, it would be useful if each end is made of a short single strand (a 
different one on each end). In both cases it would be easier to compare the 
experimental results with the theoretical predictions if artificial strands 
composed of repetitia of one base are used. Another approach would be to study 
the magnetic response of circular strands. The idea is to place many circular 
strands on a surface, apply a time-dependent magnetic flux (say, of a saw tooth 
form), and measure the response of the system. For short circular strands at 
low temperature we expect to observe persistent current carried by the charge 
solitons. Finally, a less trivial experiment, is to inject a charge soliton at 
the end of the DNA via Scanning Tunneling Microscope (STM), and look for a 
response at the other end (for example, via an attached molecule that has a 
fluorescent response to an incoming charge).

These are just three examples of many more possible tests of our predictions 
about the existence of topological charge solitons in DNA molecules. If turned 
to be verified, we expect charge solitons to have crucial role in the DNA 
activities as means of transfer of information and energy over long distances, 
and to specific locations. To study this role, one can use our model as a 
starting point, incorporating into it the inhomogeneity of the chain due to 
the different bases, as well as interactions with external molecules at 
specific sites. 

We are most thankful to E.~Braun for sharing with us his ideas about 
experimental measurements of DNA electrical transport. This research is 
supported in part by a GIF grant.


\begin{thebibliography}{10}
\bibitem{Experiment}
E.~Braun and U.~Sivan, private Communication. They have communicated with us
  their approach to measure the DNA electrical transport prior to our
  theoretical studies.
\end{thebibliography}
\end{document}